 \documentclass[prb,twocolumn,superscriptaddress]{revtex4}

\usepackage{graphicx,amsmath,color}
\usepackage{epsfig,amssymb}

\newcommand{\be}{\begin{equation}}
\newcommand{\ee}{\end{equation}}
\newcommand{\bea}{\begin{eqnarray}}
\newcommand{\eea}{\end{eqnarray}}

\newcommand{\p}{\partial}

\newcommand{\la}{\langle}
\newcommand{\ra}{\rangle}
\newcommand{\lb}{\left[}
\newcommand{\rb}{\right]}
\newcommand{\lp}{\left(}
\newcommand{\rp}{\right)}

\newcommand{\sign}{{\rm sgn}\,}
\renewcommand{\Re}{{\rm Re}\,}
\renewcommand{\Im}{{\rm Im}\,}
\renewcommand{\phi}{\varphi}
\renewcommand{\epsilon}{\varepsilon}
\renewcommand{\vec}[1]{{\bf #1}}

\begin{document}

\title{Capacitance of Graphene Bilayer as a Which-Layer Probe}
\author{Andrea F. Young}
\affiliation{Department of Physics, Columbia University, New York, NY, 10027, USA}
\author{Leonid S. Levitov}
\affiliation{Department of Physics, Massachusetts Institute of Technology, Cambridge, MA 02139, USA}

\date{\today}
\begin{abstract}
The unique capabilities of capacitance measurements in bilayer graphene enable probing of layer-specific properties that are normally
out of reach in transport measurements.
Furthermore, capacitance
measurements in the top-gate and penetration field geometries are sensitive to
different physical quantities: the penetration field capacitance probes the two
layers equally, whereas the top gate capacitance preferentially samples the near
layer,
resulting in the ``near-layer capacitance enhancement'' effect observed in recent top-gate capacitance measurements.
We present a detailed theoretical description of this effect and
show that capacitance can be used to determine the
equilibrium layer polarization, a potentially useful tool in the study
of broken symmetry states in graphene.
\end{abstract}

\maketitle
\section{Introduction}
Capacitance measurements probe the energy cost of moving charge between different parts of a system.
In a classical system, this energy cost is a purely geometric quantity and consists of the electrostatic energy. In contrast, capacitance measurements performed on quantum systems can access a range of subtle and interesting phenomena. In particular, Pauli exclusion in degenerate electronic systems gives rise to a characteristic quantum contribution to the internal energy. The associated contribution to capacitance, known as `quantum capacitance'~\cite{Luryi1988}, is proportional to the electronic compressibility $\frac{\partial n}{\partial \mu}$.
In addition, at low carrier densities, the internal energy is dominated by electronic correlations, resulting in a so-called negative compressibility contribution to capacitance~\cite{Bello1981}.
In low dimensional systems these effects can amount to a sizeable contribution, making capacitance measurements a powerful probe of many-body effects~\cite{Eisenstein1992}.  Moreover, whereas electrical transport is often dominated by a small subset of electronic states, capacitance probes all states equally. Consequently, capacitance is a useful tool in the study of
phenomena in which localization plays a role, such as quantum Hall effects and the metal-insulator transition~\cite{Eisenstein1992,Eisenstein1994,Ilani2006,Dultz2000,Ilani2001}. Under certain conditions, the quantum capacitance can become an order-one effect~\cite{Skinner2010,Li2010a}.

Graphene and its bilayer are ideal materials for the application of the capacitance technique.  The
two-dimensional geometry of these materials permits the placement
of proximal metal gates~\cite{Ponomarenko2010,Henriksen2010,Young2010}, electrolytic solutions\cite{Xia2009}, or scanning probe heads~\cite{Martin2008,Martin2009}, all of which can  be used to probe capacitance. Interesting results have been obtained for monolayer graphene, in which this quantum capacitance was found to dominate the total capacitance near the Dirac point even at room temperature~\cite{Xia2009}.
Surprisingly, the compressibility measured at low temperature~\cite{Martin2008} was found to be well described by the  noninteracting massless Dirac model,
a fact attributed to an exact cancelation of correlation effects in the monolayer\cite{Abergel2009}. In bilayer graphene (BLG), in contrast,
the interaction effects are expected to be strong, potentially leading to novel many-body states near charge neutrality\cite{Barlas2008,Vafek2009,Min2008prb2,Nandkishore2009,Lemonik2010,Jung2010,Zhang2010,Cote2010,Nandkishore2010}.
Such effects, if they exist, would directly manifest themselves
in compressibility measurements~\cite{Martin2010}.

In this paper we discuss the unique capabilities of capacitance measurements in BLG. Due to the finite interlayer separation, capacitance measurements can probe layer-specific properties that are out of reach in conventional transport measurements in which the layers are not contacted separately.  Motivated by recent experiments, we calculate the effect of a gate-induced charge imbalance between the layers on the measured capacitance in several geometries, taking into account electronic interactions and short range disorder.
We interpret the peculiar electron-hole asymmetry observed in top-gate capacitance measurements\cite{Young2010} in terms of a ``near-layer capacitance enhancement'', which is a combined effect of van Hove singularities (vHs) in the BLG band structure and the interlayer screening. We show that  capacitance experiments can be used as a which-layer probe, offering a unique capability in studying electronic properties of graphene.

\section{The near-layer capacitance enhancement}

Recently, capacitance techniques have been applied to dual-gated bilayer graphene\cite{Henriksen2010,Young2010}. The geometry of these devices allows the electrostatic potentials on the two layers to be varied independently, enabling independent control of both carrier density and the gap in the electronic spectrum ~\cite{McCann2006prl,McCann2006prb,Castro2007}.
In the absence of external fields, BLG is a metal characterized (at sufficiently low energies) by approximately parabolic valence and conduction bands which touch
at the corners of the hexagonal Brillouin zone (at the $K$ and $K'$ points).  The degeneracy
at this band crossing is protected by the symmetry of the BLG crystal structure, in which atomic sites on different layers are equivalent under transformations of the point symmetry group.  Application of an external electric field perpendicular to the layers
breaks the which-layer symmetry, turning BLG into a semiconductor with a gate-tunable band gap. At not too strong fields the gapped state can be described\cite{McCann2006prl} by projecting the tight binding Hamiltonian on the
low-energy subspace of wavefunctions $(\psi_{1}$, $\psi_{2})$ where the subscript indicates the layer index, giving the two-band Hamiltonian

\be\label{eq:H0}
H_0(\vec p)=
\left(\begin{array}{cc}
v_1&\frac{p_+^2}{2m}\\
\frac{p_-^2}{2m}&v_2
\end{array}
\right)
,\quad p_\pm=p_x\pm ip_y
,
\ee
where momentum $\vec p$ is measured relative to the $K$ (or $K'$) point and $v_1$, $v_2$ are the potentials on each layer, controlled by external gates or dopants. The Hamiltonian (\ref{eq:H0}) features a band gap of size $\Delta=|v_1-v_2|$, and a pair of vHs in the density of states of inverse square root form positioned on either side of the gap at $\epsilon=v_1$ and $\epsilon=v_2$.

The field-induced gapped state is characterized by interlayer density imbalance, in which the occupancies of the two layers are very different for $v_1=v_2$ and for $v_1\ne v_2$.
For the balanced bilayer ($v_1=v_2$) the wavefunction amplitudes on each layer are equal (up to a phase); however, in the presence of an imbalance ($v_1\ne v_2$) the amplitudes become unequal. This leads to population imbalance between the two layers,
\be\label{eq:asymmetry}
|\psi_{1(2)}(\vec p)|^2=\frac12\mp\frac12 \frac{v_1-v_2}{\sqrt{\lp p^2/m\rp^2+(v_1-v_2)^2}}
,
\ee

with a higher occupancy on the layer which has lower energy. This layer population asymmetry results in a strong asymmetry in the partial (layer specific) densities of states:
since each vHs shows up only in the partial density of states for one of the two layers, the corresponding divergent contribution to compressibility comes only from the vHs-bearing layer, remaining finite for the other layer.

\begin{figure}[t]
	\begin{center}
	\includegraphics[width=\linewidth,clip]{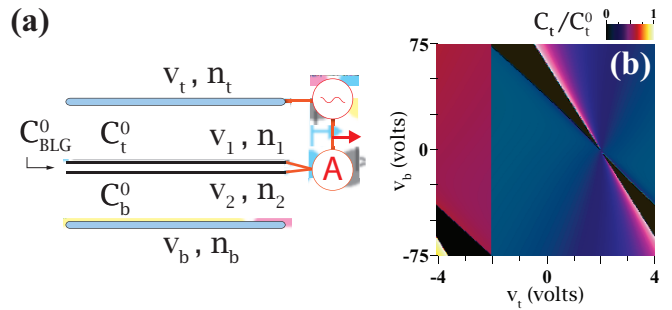}
		\caption{a)
Bilayer graphene capacitor schematic. Layer densities ($n_1$ and $n_2$) and electrostatic potentials ($v_1$ and $v_2$)
are controlled by voltages on external gates ($v_{\rm t}$ and $v_{\rm b}$), which couple
to the bilayer through the fixed geometric capacitances $C_{\rm t}^0$ and
$C_{\rm b}^0$.
Capacitance measurements~\cite{Young2010} are performed
by measuring the current flowing through \textit{both} layers in the
presence of an AC driving potential on one of the gates.
b) Top gate capacitance as a function of external gate potentials for a clean
bilayer, calculated using the self-consistent approach of Sec.\ref{sec:selfconsistent} [see Eq.(\ref{ctg}) as well as Eqs.(\ref{c1})-(\ref{c4}) and (\ref{eq:N11})-(\ref{eq:N12})]. The capacitance, which is small in the insulating regime and high in the metallic regime, is enhanced at the edges of the metallic region due to the presence of van Hove singularities in the density of states at the band edge. The enhancement is {\it asymmetric}, reflecting the asymmetric population of the layers, Eq.(\ref{eq:asymmetry}).
}
		\label{fig1}
	\end{center}
\vspace{-10mm}
\end{figure}

As we discuss in detail below, the layer population asymmetry, Eq.(\ref{eq:asymmetry}), manifests itself in capacitance measurements. This is illustrated in Fig.\ref{fig1}(b), in which top-gate capacitance found using a self-consistent model (see Sec.\ref{sec:selfconsistent}) is plotted as a function of gate voltages $V_{\rm t}$ and $V_{\rm b}$.  The enhancement in capacitance associated with the band edge is stronger when the divergent vHs-bearing layer is facing the gate used to measure capacitance (top layer for $C_{\rm t}$ and bottom layer for $C_{\rm b}$ in Fig. \ref{fig1} a). We refer to this behavior as `near-layer capacitance enhancement' (NLCE). This NLCE effect is seen in the capacitance map shown in Fig.\ref{fig1}(b): the dark region, corresponds to the insulating state realized when the chemical potential is positioned inside field-induced gap, is bordered on \textit{one} side by a bright fringe corresponding to the NLCE. The  markedly different contrast between the van
 Hove singularity- associated features positioned on either side of the dark region, is associated with the density piling up on the near layer rather than the far one.

This behavior explains the asymmetry observed in top-gate capacitance measurements [\onlinecite{Young2010}], in which a feature identified with the vHs was observed only for electrons (holes) when the high (low) energy layer was nearest the gate from which capacitance was measured.  In contrast, no such asymmetry is expected for the capacitance measured using `penetration field' geometry\cite{Henriksen2010}, because the penetration field capacitance is more symmetric than the one-sided (top or bottom) gate capacitance. Indeed, no NLCE-type asymmetry was observed in the measurements reported in Ref.[\onlinecite{Henriksen2010}]. As we shall see, the gate capacitance and the penetration field capacitance measure fundamentally different characteristics of the system. Simultaneous measurements of gate and penetration field capacitances can thus provide detailed and direct information on layer polarization of the bilayer.

The NLCE effect is sensitive to the form of the vHs, which depends on the specifics of the dispersion relation.
The simplest model for BLG, which we focus on below, is that of quartic dispersion, described by the Hamiltonian (\ref{eq:H0}). A more detailed analysis~\cite{McCann2006prl,McCann2006prb,Castro2007}, based on the four band model, leads to a `Mexican hat' structure in band dispersion near points $K$ and $K'$. However, the Mexican hat dispersion and the quartic dispersion both lead to an inverse square-root vHs at the band edge, resulting in essentially identical NLCE effects.

% The structure of this paper is as follows.
In this paper we develop theory of the NLCE effect.
In section III we calculate, using a two band model of BLG, layer-indexed densities of states, $\nu_{ij}=-\partial n_i/\partial v_j$, where $i,j,=1,2$ refer to the two layers.  In section IV we develop a many-body approach that describes interactions of particles in BLG with other particles and also with gate potentials. Using a self-consistent Hartree-type approximation, we derive expressions
for several quantities of interest relevant to capacitance measurements in terms of the matrix elements $\nu_{ij}$. We find that different experimental observables exhibit very different behavior. In particular, the gate capacitance exhibits strong particle-hole asymmetry and the NLCE effect (see Fig.1), while the penetration-field capacitance is nearly particle-hole symmetric. In section V, we consider the effect of disorder, and show that the asymmetry persists for relatively high disorder concentrations corresponding to the experimental regime.  Finally, we conclude with a discussion of the usefulness of different capacitance measurements in bilayer graphene for probing the layer-pseudospin texture of possible broken symmetry phases.

\section{The van Hove singularities and compressibility in clean BLG}

The main features of the compressibility of BLG in an external field can be understood in terms of the many-body Hamiltonian
\be\label{eq:H}
H=\sum_{\vec p,\alpha}\psi^\dagger_{\vec p,\alpha}H_0\psi_{\vec p,\alpha}
+H_{\rm int}
\ee
where $H_0$ is the single-particle Hamiltonian (\ref{eq:H0}) and summation over four flavors $\alpha=1,2,3,4$ accounts for the spin and valley ($K$, $K'$) degrees of freedom. The interaction is written in terms of density harmonics on the layers, $n_{i,\vec k}=\sum_{\vec p,\alpha}\psi_{i, \vec p,\alpha}^\dagger \psi_{i,\vec p+\vec k}$ ($i=1,2$),
\be\label{eq:Hint}
H_{\rm int}=\frac12\sum_{\vec k}\lp \begin{array}{c}n_{1,-\vec k}\\n_{2,-\vec k}\end{array} \rp^{\rm T}
\lp \begin{array}{cc}
V_{\vec k} & \tilde V_{\vec k} \\ \tilde V_{\vec k} & V_{\vec k}
\end{array} \rp
\lp \begin{array}{c}n_{1,\vec k}\\n_{2,\vec k}\end{array} \rp,
\ee
with $V_{\vec k}$ and $\tilde V_{\vec k}$ the intralayer and interlayer Coulomb interaction,
\be
V_{\vec k}=\frac{2\pi e^2}{\kappa |\vec k|}
,\quad
\tilde V_{\vec k}=e^{-|\vec k|d}V_{\vec k}
,
\ee
where $d\approx 0.3\,{\rm nm}$ is the interlayer spacing in BLG.

We analyze quantum corrections to the capacitance of gated BLG described by the Hamiltonian (\ref{eq:H}) using a Hartree-type approximation. This is done in two steps. We first find the compressibility matrix of non-interacting fermions, formally setting $H_{\rm int}=0$ in Eq.(\ref{eq:H}). In doing this, the BLG potentials $v_1$ and $v_2$ are treated as external parameters. Next, in Sec.\ref{sec:selfconsistent}, we restore the interaction $H_{\rm int}$, adding to it the interaction between all charges, including those on the gates.  We relate potentials $v_{1(2)}$ to charges on the gates and the graphene bilayer self-consistently, and use these relations to evaluate capacitance as a function of external gate voltages.

The Hartree-type analysis presented in this paper does not account for correlation effects; however, estimates of the correlation energy and the analysis of compressibility of BLG presented in Ref.[\onlinecite{Nandkishore2010}] indicate that the corresponding correction to capacitance is small, except at very low values of disorder and temperature, where the BLG system develops an instability towards a correlated state.

In recent experiments~\cite{Henriksen2010,Young2010} electronic states with different doping relative to the neutrality point are probed by varying the potentials $v_1$ and $v_2$ through their response to the potentials $v_{\rm t}$ and $v_{\rm b}$ applied to external gates. Metallic and insulating conductance regimes occur when the Fermi level lies inside or outside the gate-induced gap~\cite{Oostinga2008,Zou2010,Taychatanapat2010}.  The insulating regime was observed to accompany a drop in compressibility.

It is convenient to introduce layer-symmetrized potentials $v_\pm=\frac12(v_1\pm v_2)$. Within the two band model (\ref{eq:H0}), the gap size is $\Delta=2|v_-|$ and the position of the gap center relative to the Fermi level is $v_+-\mu$; the metallic and insulating regimes in a clean bilayer are then described by $|v_+ -\mu|>|v_-|$ and  $|v_+ -\mu|<|v_-|$, respectively. In experiments~\cite{Henriksen2010,Young2010} capacitance was measured with the graphene bilayer grounded.  This situation can be described by a Fermi level pinned to zero energy, $\mu=0$.

Particle densities on the two layers can be expressed as sums over all occupied states,
% with $\epsilon<\mu=0$ as
%
\be
n_{1(2)}=\int \frac{d^2 p}{(2\pi\hbar)^2} f(\vec p) |\psi_{1(2)}(\vec p)|^2
% \frac{1}{V}\sum_{p,\epsilon(\vec p)<0}|\psi_{1(2)}(\vec p)|^2
,
\ee
where $f(\vec p)=1/(e^{\beta\epsilon(\vec p)}+1)$. In what follows, we focus on the case of zero temperature, $f(\vec p)=\theta(-\epsilon(\vec p))$.
%Below it will be convenient to use densities with an added minus sign, $\tilde n_{1(2)}=- n_{1(2)}$. The minus sign ensures that positive $v_{1(2)}$ produces positive density, leading to positive capacitance in Sec.\ref{sec:selfconsistent}.
%
Using the eigenstates of the Hamiltonian (\ref{eq:H0}) and defining layer-symmetrized densities $n_\pm=n_1\pm n_2$, we find
% $v_\pm=\frac12(v_1\pm v_2)$, we find
% \footnotemark
% Straightforward calculation gives
%
\bea\label{eq:n_+}
&& n_+=\begin{cases}
-\nu_0\sqrt{v_+^2-v_-^2}\, \sign v_+  & $(metal),$
%|v_+|>|v_-|
\\ 0 & $(insulator),$
%|v_+|<|v_-|
\end{cases}
%=\nu_0\sqrt{v_1v_2}\, \sign (v_1+v_2)
%,\quad
\\\label{eq:n_-}
&& n_-=\begin{cases}
-\nu_0 v_-\ln \lp \frac{2\Lambda}{|v_+|+\sqrt{v_+^2-v_-^2}}\rp  & $(metal),$
% |v_+|>|v_-|
\\
-\nu_0 v_-\ln \lp \frac{ 2\Lambda}{|v_-|}\rp  & $(insulator),$
% |v_+|<|v_-|
\end{cases}
%=\frac{\nu_0(v_1-v_2)}2 \ln \lp \frac{4\Lambda}{|v_1+v_2|+2\sqrt{v_1v_2}}\rp
% ,\quad
% \nu_0=\frac{2m e^2}{\pi\hbar^2}
% n_{1(2)}=\sum_{\vec p}\frac12\lp 1\pm\cos\theta_p\rp
\eea
where $\Lambda$ is an ultraviolet cutoff of order the bandwidth. Here $\nu_0=2m e^2/(\pi\hbar^2)$ accounts for the four-fold spin/valley degeneracy, and can be written as $2/\pi a_B$, where $a_B$ is the Bohr's radius of BLG. The two cases in Eqs.(\ref{eq:n_+}),(\ref{eq:n_-}), metallic and insulating,  correspond to the regimes $|v_+|>|v_-|$ and $|v_+|<|v_-|$.

\begin{figure}[t]
	\begin{center}
	\includegraphics[width=\linewidth,clip]{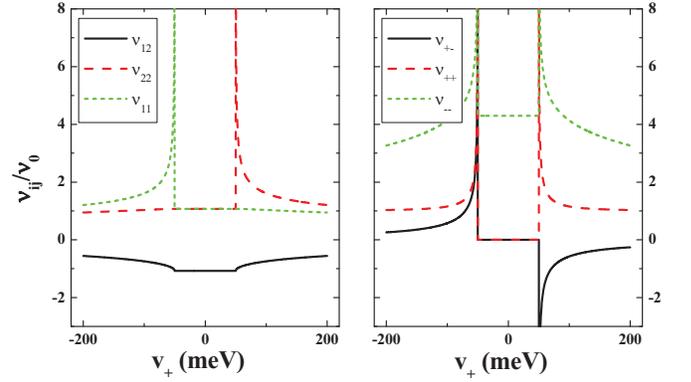}
		\caption{Energy dependence of the interlayer compressibility matrix
elements $\nu_{ij}$ in the 1/2 (left panel, Eqs.
(\ref{eq:N11})-(\ref{eq:N12})) and $\nu_{\pm}$ (right panel, Eqs.
(\ref{eq:nupp})-(\ref{eq:nupm})) bases for fixed interlayer asymmetry $v_- =50{\rm meV}$
and $\Lambda=5{\rm eV}$.  In the left panel, single layer charge
compressibilities $\nu_{11}$ and $\nu_{22}$ are divergent only on one side
of the charge gap, allowing the interlayer asymmetry to be probed by single
side capacitance measurements. In the $+/-$ basis, this asymmetry is reflected
by the charge-flavor response, $\nu_{+-}$.}
		\label{fig3}
	\end{center}
\end{figure}

Using these expressions we can compute the entries of the compressibility matrix $\nu_{ij}=-\p n_i/\p v_j$. The expressions have different form for  $|v_+|>|v_-|$ and for  $|v_+|<|v_-|$:
\bea\label{eq:nupp}
&&\nu_{++}=\begin{cases}\nu_0 \frac{|v_+|}{\sqrt{v_+^2-v_-^2}}  & $(metal),$
% |v_+|>|v_-|
\\ 0 &  $(insulator),$
% |v_+|<|v_-|
\end{cases}
\\
&&\nu_{--}=\begin{cases}\tilde\nu_0 +\nu_0
\frac{|v_+|}{\sqrt{v_+^2-v_-^2}} & $(metal),$
% |v_+|>|v_-|
\\
\nu_0 \ln \lp \frac{2\Lambda}{e |v_-|}\rp & $(insulator),$
% |v_+|<|v_-|
\end{cases}
% |v_-|}-\cosh^{-1}\frac{|v_+|}{|v_-|}\rp
\\\label{eq:nupm}
&&\nu_{+-}=\nu_{-+}=\begin{cases}
-\nu_0 \frac{v_- \sign v_+}{\sqrt{v_+^2-v_-^2}} & $(metal),$
% |v_+|>|v_-|
\\ 0 & $(insulator),$
% |v_+|<|v_-|
\end{cases}
% \\
% &&\p \tilde n_-/\p v_+=-\nu_0 \frac{v_- \sign v_+}{\sqrt{v_+^2-v_-^2}},
\eea
where we defined
\be
\tilde\nu_0 =\nu_0 \ln \lp \frac{2\Lambda}{ e\lp |v_+|+\sqrt{v_+^2-v_-^2}\rp } \rp
,
\ee
with $e=2.71828...$. Expressions (\ref{eq:nupp})-(\ref{eq:nupm}) are plotted in the left panel of Fig. \ref{fig3}.  Note that the compressibility matrix is symmetric, $\nu_{+-}=\nu_{-+}$.

Different elements of matrix $\hat \nu$ have different physical meanings. The diagonal element  $\nu_{++}=-\partial n_+/\partial v_+$ is the total charge compressibility. The diagonal element $\nu_{--}=-\partial n_-/\partial v_-$ is layer polarizability. The off-diagonal elements $\nu_{-+}=\nu_{+-}=-\partial n_-/\partial v_+$ describe the charge-flavor response.
The latter quantities are particularly useful, as they measure the layer distribution of incremental additions of charge, giving information about the layer polarization of the ground state: the quantities $\nu_{-+}$ and $\nu_{+-}$ are zero for an unpolarized bilayer, but nonzero in the presence of a charge imbalance.

Rewriting Eqs.(\ref{eq:n_+}),(\ref{eq:n_-}) in terms of variables characterizing individual layers, $n_1$, $n_2$, we obtain
% and $v_1$, $v_2$, we obtain
%
\bea\label{eq:N11}
&&\nu_{11}=\frac12\nu_0\frac{|v_+|-v_-\sign v_+}{\sqrt{v_+^2-v_-^2}}
+\frac14\tilde\nu_0
\\\label{eq:N22}
&&\nu_{22}=\frac12\nu_0\frac{|v_+|+v_-\sign v_+}{\sqrt{v_+^2-v_-^2}}
+\frac14\tilde\nu_0
\\\label{eq:N12}
&&\nu_{12}=\nu_{21}=
% \frac{\p \tilde n_2}{\p v_1}=\frac14\nu_0\frac{|v_+|-v_+}{\sqrt{v_+^2-v_-^2}}
-\frac14\tilde\nu_0
.
\eea
%
% where $\tilde\nu_0=\nu_0\ln\lp 2\Lambda/\lp e\lp |v_+|+\sqrt{v_+^2-v_-^2}\rp\rp\rp$.
Expressions (\ref{eq:N11})-(\ref{eq:N12}) are invariant under simultaneous $1\leftrightarrow 2$ exchange and gap inversion, $v_-\rightarrow -v_-$.

Both of the diagonal compressibility matrix elements ($\nu_{11}$ and $\nu_{22}$) exhibit an inverse square root divergence at the charge gap edge, where the density of single particle states has a van Hove singularity. The two diagonal compressibilities behave {\it asymmetrically}, diverging on opposite sides of the gap:
$\p n_1/\p v_1$ diverges at $v_1\to 0$, while $\p n_2/\p v_2$ diverges at $v_2\to 0$. In contrast, the off-diagonal compressibilities ($i\ne j$) remain finite on either side of the charge gap and are symmetric (see Fig. \ref{fig3}, left panel). Inside the charge gap, $|v_+|<|v_-|$, the diagonal and off-diagonal compressibilities are constant:
\be
\nu_{11}=\nu_{22}=-\nu_{12}=-\nu_{21}=\frac{\nu_0}4 \ln \lp \frac{2\Lambda}{e |v_-|}\rp
% (\tilde\nu_0-\nu_0)
% \\
% && \frac{\p \tilde n_1}{\p v_2}=-\frac14\nu_0 \ln \lp \frac{2\Lambda}{e |v_-|}\rp
% (\tilde\nu_0-\nu_0)
,
\ee
exhibiting no divergence at the gap edge.

% \footnotetext{
% Eqs.(\ref{eq:n_pm_answers}) are valid for $|v_+|>|v_-|$, the cndition that can also be written as $v_1v_2>0$. For $|v_+|\le |v_-|$, or $v_1v_2\le 0$, we have
% $n_+=0$, $n_-=\nu_0 v_-\ln\frac{2\Lambda}{|v_-|}$.
% }

\section{Self-consistent capacitance calculation}
%{Electrostatic analysis}
\label{sec:selfconsistent}

We shall focus on the geometry pictured in Fig.\ref{fig1}a, which describes a dual-gated graphene device of the type studied in Refs.[\onlinecite{Young2010}] and [\onlinecite{Henriksen2010}].  The experimental system consists of a bilayer graphene sheet placed between two gates, characterized by potentials $v_{\rm t}$ and $v_{\rm b}$, charge densities $n_{\rm t}$ and $n_{\rm b}$, and geometric capacitances to the bilayer $C_{\rm t}^0$ and $C_{\rm b}^0$. The bilayer is described by the potentials $v_1$ and $v_2$ and charge densities $n_1$ and $n_2$ induced by the external gates on the individual layers.  Electrostatic energy of the bilayer itself is taken into account by including an interlayer capacitance $C_{\rm BLG}$, which can be estimated from the ``geometric'' value obtained for a parallel plate capacitor, $C_{\rm BLG}=(4\pi d)^{-1}$, with $d\approx 0.3\,{\rm nm}$.  This electrostatic model amounts to the approximation that the charge density on the bilayer is of the for
 m $n(z)=n_1\delta(z-d/2)+n_2\delta(z+d/2)$. While corrections are expected due to the finite extent of the wavefunctions, these corrections amount, for the most part, to a renormalization of $C_{\rm BLG}$, upon which our results do not sensitively depend.

The quantities of interest obey the general electrostatic charge field relations
\bea
&& C_{\rm t}^0(v_{\rm t}-v_1)=\frac12 \lp n_{\rm t}-n_1-n_2-n_{\rm b}\rp,\label{c1}
\\
&& C_{\rm BLG}^0(v_1-v_2)=\frac12\lp  n_{\rm t}+n_1-n_2-n_{\rm b}\rp,
\\
&& C_{\rm b}^0(v_2-v_{\rm b})=\frac12\lp  n_{\rm t}+n_1+n_2-n_{\rm b}\rp,
\\
&& n_{\rm t}+n_1+n_2+n_{\rm b}=0 \label{c4}
.
\eea
To complete the system of equations for charge densities and potentials, a set of constitutive relations for BLG must be used. These relations, which are of general form $n_1=f_1(v_1,v_2)$, $n_2=f_2(v_1,v_2)$, will be calculated in subsequent sections.

Capacitance measurements are done in the finite frequency regime, by applying a small AC bias (on top of the DC bias used to control density and interlayer imbalance) to one terminal of the device and then recording the resulting change in charge density on a second terminal.  Choice of terminals distinguishes top (back) gate capacitance, $C_{\rm t(b)}$, from penetration field capacitance, $C_{\rm p}$,
\be
C_{\rm t(b)}=-\left.\frac{\delta n_1+\delta n_2}{\delta v_{\rm t(b)}}\right|_{\delta v_{\rm b(t)}=0};\quad C_{\rm p}=- \left.\frac{\delta n_{\rm t}}{\delta v_{\rm b}}\right|_{\delta v_{\rm t}=0}.
\ee
After eliminating  $n_{\rm t}$ and $n_{\rm b}$ from Eqs. (\ref{c1})-(\ref{c4}) by expressing them in terms of other variables, $n_{\rm t}=C_{\rm t}^0(v_{\rm t}-v_1)$, $n_{\rm b}=C_{\rm b}^0(v_{\rm b}-v_2)$, the remaining two equations are linearized with the help of the matrix of inter- and intralayer compressibilities
\be
\hat \nu = -\lp \begin{array}{cc}
\frac{\p  n_1}{\p v_1} & \frac{\p  n_1}{\p v_2} \\ \frac{\p  n_2}{\p v_1} & \frac{\p  n_2}{\p v_2}
\end{array} \rp,\quad
\lp \begin{array}{c}
\delta n_1 \\ \delta n_2
\end{array} \rp
=
-\hat \nu \lp \begin{array}{c}
\delta v_1 \\ \delta v_2
\end{array} \rp
.
\ee
This yields
\be
\lb
\hat \nu
+\hat C
\rb
\lp \begin{array}{c}
\delta v_1 \\ \delta v_2
\end{array} \rp
=
\lp \begin{array}{c}
 C_{\rm t}^0 \delta v_{\rm t}\\ C_{\rm b}^0 \delta v_{\rm b}
\end{array} \rp
\ee
where $\hat C$ is a matrix of geometric capacitances,
\be
\hat C=\lp \begin{array}{cc}
C_{\rm BLG}^0+C_{\rm t}^0 & -C_{\rm BLG}^0 \\ -C_{\rm BLG}^0 & C_{\rm BLG}^0+C_{\rm b}^0
\end{array}
\rp
.
\ee
These expressions account for both the geometric and `intrinsic' capacitance of BLG.

Solving for $\delta v_{1}$, $\delta v_{2}$, we find the charges induced on each layer by the gate potentials:
\bea\label{eq:n_12}
&&\lp \begin{array}{c} \delta n_1 \\ \delta n_2 \end{array} \rp
= \lb \hat 1 -\hat C\lp
\hat \nu
+
\hat C \rp^{-1}\rb
\lp \begin{array}{c}
 C_{\rm t}^0 \delta v_{\rm t}\\ C_{\rm b}^0 \delta v_{\rm b}
\end{array} \rp
\eea
Here the first term describes the geometric capacitance, which would be the only contribution if the electronic system in BLG was infinitely compressible, $\hat \nu\to\infty$. The term proportional to $-\hat C(\hat \nu+\hat C)^{-1}$ describes the quantum capacitance contribution.  Combining equation (\ref{eq:n_12}) with the relations for $n_{\rm t}$ and $n_{\rm b}$, all three capacitance observables can be calculated:
\begin{align}
C_{\rm t}&=C_{\rm t}^0\left(1-\frac{\det(\hat C)-C_{\rm b}^0 \nu_{21}+C_{\rm t}^0 \nu_{22}}{\det (\hat\nu+\hat C)}\right)\label{ctg}\\
C_{\rm b}&=C_{\rm b}^0\left(1-\frac{\det(\hat C)-C_{\rm t}^0 \nu_{12}+C_{\rm b}^0 \nu_{11}}{\det (\hat\nu+\hat C)}\right)\label{cbg}\\
C_{\rm p}&=\frac{C_{\rm b}^0 C_{\rm t}^0}{\det (\hat \nu+\hat C)}\left(C_{\rm BLG}^0-\nu_{21}\right). \label{cpg}
\end{align}
These quantities implicitly depend on the gate potentials through the compressibility matrix $\nu_{ij}$.

Notably, different capacitance observables depend on different combinations of the compressibility matrix elements, and
obey different symmetries. The penetration field capacitance $C_{\rm p}$ is dominated by the off diagonal component of the (necessarily symmetric) compressibility matrix. As a result, for a symmetric device ($C_{\rm b}^0=C_{\rm t}^0$) it is invariant under interchanging layers 1 and 2 and therefore does not exhibit the NLCE effect.
%, so that for a symmetric device ($C_{\rm b}^0=C_{\rm t}^0$) it is invariant under interchanging layers 1 and 2 and does not manifest signatures of the NLCE.
In contrast, the expressions for $C_{\rm b}$ and $C_{\rm t}$
%, in contrast,
are not  $1\leftrightarrow2$ invariant.
In particular, the last term in the expression for $C_{\rm t}$,
% for example,
proportional to $\nu_{22}$, changes to $\nu_{11}$ upon layer permutation.
%its layer-conjugate $\nu_{11}$.
As shown in the previous section, in the presence of a layer imbalance these two quantities are not the same, leading to the observed NLCE observed in Ref.\onlinecite{Young2010}.

\begin{figure}[t]
	\begin{center}
	\includegraphics[width=\linewidth,clip]{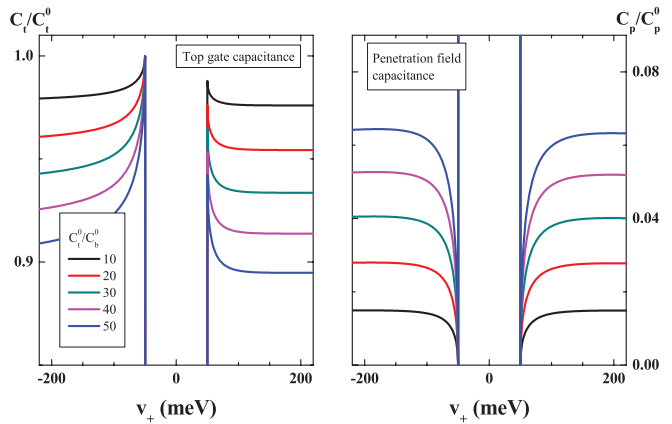}
\caption{Calculated $C_{\rm t}$ (a) and $C_{\rm p}$ (b) for the clean bilayer. Different color traces correspond to different values of the top gate capacitance, measured relative to a fixed $C_{\rm b}^0$ (taken to be 120 aF/$\mu{\rm m}^2$ corresponding to the standard 285 nm SiO$_2$). Penetration field traces are normalized by the geometric value corresponding to full penetration, $C_{\rm p}^0=\lp 1/C_{\rm b}^0+1/C_{\rm BLG}^0+1/C_{\rm t}^0\rp^{-1}$.}
%Calculated $C_{\rm t}$ (a) and $C_{\rm p}$ (b) for the clean bilayer. Different color traces correspond to different values of the top gate capacitance, measured relative to a fixed $C_{\rm b}^0$ (taken to be 120 aF/$\mu{\rm m}^2$ corresponding to the standard 285 nm SiO$_2$).  Penetration field traces are normalized by the full penetration value, $C_{\rm p}^0=\lp 1/C_{\rm b}^0+1/C_{\rm BLG}^0+1/C_{\rm t}^0\rp^{-1}$.}
		\label{fig2}
	\end{center}
\end{figure}

In a device in which all capacitances can be measured, combinations of the measured quantities can be combined to probe the charge-flavor response.  For the simplest case of a symmetric gate configuration ($C_{\rm b}^0=C_{\rm t}^0$), \begin{equation}
  \frac{C_{\rm t}-C_{\rm b}}{C_{\rm p}}=\frac{4 \nu_{-+}}{4 C_{\rm BLG}^0+\nu_{--}-\nu_{++}}.
\end{equation}
Because this quantity is proportional to $\nu_{-+}$, it can be used to probe both gate-induced and spontaneous layer polarization, allowing direct experimental measurement---somewhat analogous to Knight Shift measurements for spin---of the ground state layer polarization.

\section{The effect of disorder}
In the devices used for capacitance measurements in Refs.[\onlinecite{Young2010}],[\onlinecite{Henriksen2010}], graphene flakes were supported by a silica substrate. The carrier mobility in such devices was of order 1,000 cm$^2$/V sec. For such low-mobility devices, taking into account the effect of disorder is crucial for developing a sensible model of the experimental data. Full quantitative description of experiments requires including realistic disorder, which is likely long range~\cite{Nilsson2007,Nilsson2008,Abergel2011}, along with the effects of electronic correlations~\cite{Borghi2010} which can give quantitative corrections to the electronic compressibility.  However, the the key features of the data are captured by a simpler short range disorder model~\cite{Mkhitaryan2008}, which involves delta-function impurities localized on carbon sites:

\be\label{eq:hamiltonian1}
H=\sum_{\vec p} \psi^{\dagger}_{\bf p}H_0\psi_{\bf p}+\sum_{\bf x} u({\bf x})\psi^{\dagger}_{\bf x} \psi_{\bf x}
% ,\quad u(\vec x)=\sum_i U\delta(\vec x-\vec x_i)
,
\ee
with potential $u(\vec x)=\sum_i U\delta(\vec x-\vec x_i)$ taking values $U$ on the carbon sites occupied by impurities, and zero elsewhere. The impurities are assumed to be distributed randomly with concentration $n$.

% An analytic approach to analyze transport properties, such as disorder scattering and conductivity,
The problem (\ref{eq:hamiltonian1}) can be analyzed using a self-consistent T-matrix approximation (SCTA). The SCTA approach provides a somewhat more general approach than the self-consistent Born approximation, and is reduced to the latter for weak disorder.
% Next, we introduce an analytical mean-field approach, and employ it to analyze the DOS and transport properties of the system.
% In the model (\ref{eq:hamiltonian1}), the $T$-matrix of an individual adatom has a resonant form~\cite{Pepin01},
% \be\label{eq:T}
% T_0(\epsilon)=\frac{\pi v_0^2}{\epsilon \ln (iW/\epsilon)+\delta}
% ,\quad \delta=\frac{\pi v_0^2}{\tilde U}
% ,\quad \tilde U=U/n_0
% ,
% \ee
% where $W\approx 3t_0$ is the bandwidth, and the parameter $\delta$ determines the energy of the resonance, $\epsilon_0 \ln W/|\epsilon_0|=-\delta$. We note that the description of adatoms by an on-site potential is equivalent to the models \cite{Wehling08,Falko08}, which describe adatom states in terms of a localized level hybridized with the graphene continuum, since the form of the T-matrix at low-energies is the same in both approaches.

% {\bf changes start here..}
We evaluate the DOS and the total energy by employing disorder-averaged Greens functions
%obtained from a self-consistent $T$-matrix approximation (SCTA). We use the Greens function
expressed through the layer-indexed disorder-averaged self-energies $\Sigma_i$
% \mpar{LL: changed sign of $t_{\vec p}$}
%
\be\label{eq:Gmatrix}
G(\epsilon,\vec p)=\left[\begin{array}{cc}
         \epsilon-v_1-\Sigma_1  & -t_{\vec p}  \\
         -t_{\vec p}^*&  \epsilon-v_2-\Sigma_2     \end{array}
\right]^{-1}
% , \quad
% \epsilon_{1(2)}=\epsilon-\Sigma_{1(2)}(\epsilon)
,
\ee
%
%with $t_{\vec k}=t_0(1+e^{-i{\vec k \vec e_1}}+e^{-i{\vec k\vec e_2}})$.
 where t$_{\bf p}$ is the kinetic energy operator~\cite{McCann2006prl,McCann2006prb,Castro2007}, $t_{\vec k}\propto (1+e^{-i{\vec k \vec e_1}}+e^{-i{\vec k\vec e_2}})^2$.  An infinitesimal imaginary part $\pm i0$ should be added to $\epsilon$ to obtain the retarded and advanced Greens functions.

\begin{figure}
\includegraphics[width=\linewidth]{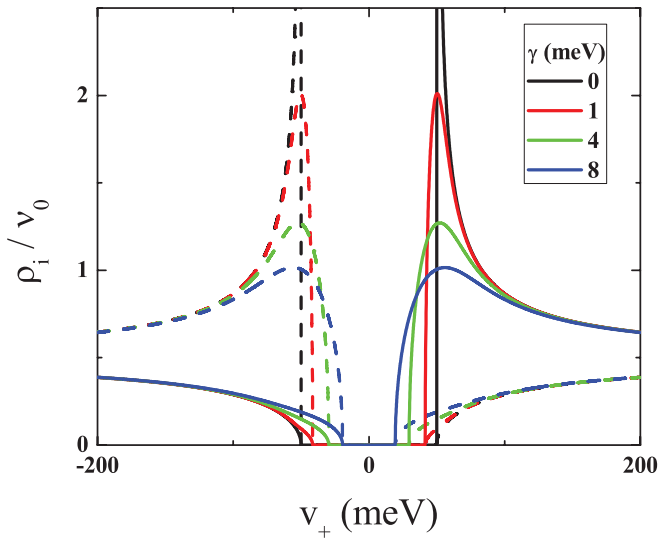}
\vspace{-4mm}
\caption[]{The effect of disorder on the density of states. Partial density of states $\rho_i$, Eq. (\ref{eq:rho}) for layers $i=1$ (solid lines) and $i=2$ (dashed lines) of a graphene bilayer, obtained from the self-consistent Born approximation, Eqs.(\ref{eq:DOS}),(\ref{eq:lambda}). Increasing the disorder strength leads to smearing of van Hove singularities and, eventually a closing of the energy gap.
}
\vspace{-4mm}
\label{fig2a}
\end{figure}

The self-energy is approximated by the average values of the $T$-matrix, evaluated separately for the sites on layers $1$ and $2$,
\be\label{eq:self-energy}
% \Sigma_{1(2)}(\epsilon)=n_{1(2)}^*\la T(\epsilon)\ra _{1(2)}
\Sigma_1 (\epsilon)=\tilde n\la T_1 (\epsilon)\ra
,\quad
\Sigma_2 (\epsilon)=\tilde n\la T_2 (\epsilon)\ra
% \Sigma_{1(2)}(\epsilon)=n_{1(2)}^*T(\epsilon),
.
\ee
Here $\tilde n=n \rho_0$ is the adatom density with $\rho_0=2/3\sqrt{3}a^2$ the density of type $1$ sites. The quantities $T_{1(2)}$, written as a $2\times 2$ matrix, are given by
\be\label{eq:Tmatrix}
% T(\epsilon)=
\left[\begin{array}{cc}
         T_1  & 0  \\
         0&  T_2      \end{array}
\right]=\frac{\tilde{U}}{1-\tilde{U} g}, \,\, g=\int \frac{d^2p}{(2\pi)^2} G(\epsilon,\vec p)
,
\ee
where $\tilde U=U/\rho_0$.
For realistic values of $v_1$ and $v_2$ the integral of the Greens function over the Brillouin zone is dominated by the regions near $K$ and $K'$; approximating $t_{\vec p}\approx (p_x\pm ip_y)^2/2m$, we obtain
%
% In the above equation, the integration is performed near one of the corners $K,K'$ of the Brillouin zone, and the angular integration makes the matrix structure of $g$ trivial. As a next step, we express $g$ in terms of $\epsilon_1 , \epsilon_2$~\cite{Shytov09},
%
\be\label{eq:g}
g=\frac{-i m}{2\sqrt{\epsilon_1\epsilon_2}}
%-\frac{\log (-W^2/\epsilon_1 \epsilon_2)}{2\pi v_0^2}
\left[\begin{array}{cc}
       {\epsilon_2} & 0  \\
         0 &  {\epsilon_1}
      \end{array}
\right]
,\quad
\epsilon_{1(2)}=\epsilon-v_{1(2)}-\Sigma_{1(2)}(\epsilon)
.
\ee
This expression is valid for $\epsilon_{1(2)}$ small compared to the bandwidth. Combining this result with Eq.(\ref{eq:self-energy}), we obtain two coupled equations for $\epsilon_1$, $\epsilon_2$:
\be\label{eq:eAeB}
\epsilon_1 =\epsilon-v_1-\frac{n U}{1+i\beta/ \lambda(\epsilon) }, \,\,\epsilon_2 =\epsilon-v_2-\frac{n U}{1+i\beta \lambda(\epsilon) },
\ee
where we defined $\lambda(\epsilon)=\sqrt{\epsilon_1 /\epsilon_2}$ and $\beta= m\tilde U/2$.
% used the relation $U=\tilde{U}n_0$, and defined
% \be\label{eq:lambda}
% \gamma= \frac{mU}2
% \frac{1}{\sqrt 3\pi} \frac{ U}{t_0^2} \log\left( - \frac{W^2}{\epsilon_1 \epsilon_2}\right)
% .
% \ee
%
Solving these equations
% numerically
for $\epsilon_1$, $\epsilon_2$ as a function of $\epsilon$, we find the Greens function (\ref{eq:Gmatrix}) and use it to calculate the density of states,
\be\label{eq:DOS}
\rho(\epsilon)=\frac{1}{\pi}\Im\!\!\int G(\epsilon+i0,\vec p)\frac{d^2p}{(2\pi)^2}=\frac{m}{\pi}
%-\frac{\log (-W^2/\epsilon_1 \epsilon_2)}{2\pi v_0^2}
\left[\begin{array}{cc}
       \lambda^{-1}(\epsilon) & 0  \\
         0 &  \lambda(\epsilon)
      \end{array}
\right]
,
%\frac{1}{2\pi^2 v_0^2} {\rm Im  }\left[ -i(\epsilon_1+\epsilon_2) \log\frac{W^2}{\epsilon_1 \epsilon_2} \right]  %|_{i\epsilon\to \epsilon+i0}.
\ee
where the integral is identical to the one in Eq.(\ref{eq:g}). A factor of two was inserted after integration to account for spin degeneracy.

The density of states is expressed through the quantity $\lambda(\epsilon)$. Taking the ratio of the self-consistent equations for $\epsilon_1$ and  $\epsilon_2$, Eq.(\ref{eq:eAeB}), we obtain a single equation for the quantity $\lambda$. Focusing on the case of weak disorder potential and expanding in $U$, we arrive at
\be\label{eq:lambda}
\lambda^2=\frac{\epsilon-v_1 +i \gamma/\lambda}{\epsilon-v_2 +i \gamma\lambda}
,\quad
\gamma=\frac{mU^2}{2\rho_0}n
,
\ee
where the terms linear in $U$ have been incorporated in the quantities $v_{1(2)}$. Once $\lambda(\epsilon)$ is found from Eq.(\ref{eq:lambda}), it can be plugged into Eq.(\ref{eq:DOS}) to obtain partial densities of states on each of the layers (see Fig.\ref{fig2}),
\be\label{eq:rho}
\rho_1(\epsilon)=\frac{\nu_0}{2}\Re \lambda^{-1}
,\quad
\rho_2(\epsilon)=\frac{\nu_0}{2}\Re \lambda
.
\ee
In the absence of disorder, $\gamma=0$, we have $\lambda=\sqrt{(\epsilon-v_1)/(\epsilon-v_2)}$, which gives van Hove singularities of an inverse square root form at the band edges $\epsilon=v_1,v_2$ as found in section I. In the presence of disorder, these singularities are washed out to varying degrees.  As shown in Fig.\ref{fig2a}, this washing out proceeds by both reducing the height of the vHs peak and closing the gap.  Crucially, the `off'-layer density of states at the energy of the `on' layer vHs peak increases with disorder.  This has the effect of increasing the screening effect of the `off' layer when it lies closer to the gate used to measure capacitance, \textit{enhancing} the NLCE effect for disordered samples.

\begin{figure}[!]
	\begin{center}
	\includegraphics[width=\linewidth]{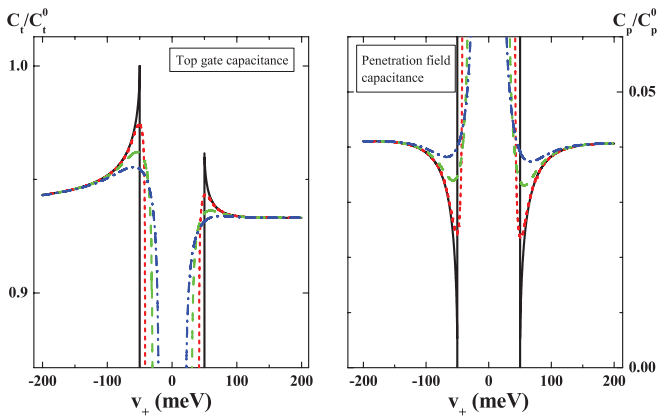}
		\caption{Top gate (left panel) and penetration field (right panel)
capacitance for different values of the short-range disorder parameter
$\gamma$, here measured in meV.  Interlayer asymmetry parameter $v_- =50{\rm meV}$ and the cutoff
$\Lambda=5{\rm eV}$.  Geometric parameters are chosen to match experiment
reported in Ref.[\onlinecite{Young2010}], $C_{\rm t}^0/C_{\rm b}^0 =30$, $C_{\rm b}^0=120{\rm aF/\mu m}^2$.  Color scheme corresponds to varying values of $\gamma$ as in Fig. \ref{fig2a}.}
		\label{fig4}
	\end{center}
\end{figure}

To calculate experimental capacitances, Eqs. (\ref{cbg})-(\ref{cpg}), the partial densities of states are integrated numerically with respect to energy and then redifferentiated with respect to the appropriate energy variable, $v_1$ or $v_2$. In Figure \ref{fig4}, the results for both top gate and penetration field capacitance for a device with electrostatic parameters resembling those in Ref.\onlinecite{Young2010} are plotted.  The asymmetry of top gate capacitance survives disorder averaging, and indeed is enhanced.  For intermediate values of disorder, electrons and holes display qualitatively different behavior: the non-monotonic vHs feature survives for holes but is completely obliterated for electrons, as observed in Ref.\onlinecite{Young2010}.

\section{Conclusions}

As we argue above, electrostatic capacitance measurements offer a unique which-layer probe for BLG. The sensitivity to the interlayer imbalance arises despite the fact that the layers are not contacted separately: the relative proximity of the layers to the top- and bottom- gates, combined with the interlayer screening, allows capacitance measurements to access layer specific quantities.  Gate capacitance measurements preferentially probe the nearer layer, leading to the NLCE effect as the near layer screens the far layer.  Consequently, in the presence of a layer imbalance, top- and bottom- gate capacitance measurements will be different. This difference is the signature of layer polarization, allowing its unambiguous experimental determination.

Our analysis provides an explanation of
% Most of this paper was devoted to explaining
recent top gate capacitance experiments on dual gate bilayer graphene structures~\cite{Henriksen2010,Young2010}.
% in terms of the bilayer band structure, disorder, and the NLCE effect.
Since the degeneracy of the band crossing in the BLG spectrum at the $K$ and $K'$ points is linked to inversion symmetry, the gate-induced density imbalance and the opening of a band gap go hand in hand~\cite{McCann2006prl,McCann2006prb,Castro2007}.
As we have shown, this imbalance can be probed directly through NLCE measurements;
to our knowledge, the NLCE-type asymmetry observed in Ref.\onlinecite{Young2010}
is the first direct experimental evidence of layer imbalance in BLG.

The possibility of probing layer polarization directly through capacitance measurements has implications
beyond the study of gate-induced gap opening.  Recently, experimental sample quality has improved to the point of allowing the observation of a multitude of novel features likely associated with electronic correlations~\cite{Feldman2009,Zhao2010,Dean2010,Martin2010,Weitz2010}. A large number of possible broken symmetry states, arising in the presence and in the absence of magnetic field, have been explored in the theoretical literature~\cite{Barlas2008,Vafek2009,Min2008prb2,Nandkishore2009,Lemonik2010,Jung2010,Zhang2010,Cote2010,Nandkishore2010}, including several mutually exclusive scenarios for the ordering at low densities and small electric and magnetic fields. The main open questions pertaining to these states have to do with identifying broken symmetries and determining the exact structure of the order parameter and excitations.
%
%LL In the presence of quantizing magnetic field, the combined spin and sublattice symmetry produce multiply degenerate Landau levels in BLG electronic spectrum.
%LL1 four-fold degenerate Landau levels; in the the zeroth Landau level, an additional orbital degeneracy due to the $2\pi$ Berry phase adds an additional orbital spin degree of freedom, bringing to the total degeneracy to eight \cite{Novoselov2006,McCann2006prl,Castro2007}.
Future NLCE measurements, by offering a direct method for determination of the layer polarization, will help to narrow down the possibilities for these new states.

\acknowledgements

We thank P. Kim and R. Nandkishore for useful discussions.
This work was supported by Office of Naval Research Grant
No. N00014-09-1-0724 and the Department of Energy under DOE (DE-FG02-05ER46215).

\end{document}